\documentclass[aps,pre,10pt,showpacs,superscriptaddress,reprint]{revtex4-1}

\usepackage{times}
\usepackage{amsmath}
\usepackage{amssymb}
\usepackage{amsfonts}
\usepackage[dvips]{graphicx,color}
\usepackage{epsfig}
\usepackage{color}

\newcommand{\rf}[1]{~(\ref{#1})}
\newcommand\f[1]{~Fig.~\ref{fig:#1}}

\def\dst{\displaystyle}
\def\sst{\scriptstyle}

\def\Zef{Z_{\ast}}

\newcommand{\chir}{\chi_{_{\!\!\sst{Z_1}}}}
\newcommand{\chii}{\chi_{_{\!\!\sst{Z_2}}}}
\newcommand{\chiii}{\chi_{_{\!\!\sst{Z_3}}}}

\newcommand{\alphaz}{\alpha_{_{\!\!\sst{Z}}}}
\newcommand{\betaz}{\beta_{_{\!\!\sst{Z}}}}
\newcommand{\betazz}{\widetilde{\beta}_{_{\!\!\sst{Z}}}}
\newcommand{\fF}{f_{\sst\mathcal{F}}}
\newcommand{\emu}{\varepsilon_{\mu}}
\newcommand{\eF}{\varepsilon_{\sst\mathrm{F}}}

\begin{document}

\title{Dielectric function of a collisional plasma for arbitrary ionic charge}

\author{H.~B.~Nersisyan}
\email{hrachya@irphe.am}
\affiliation{Theoretical Physics Division, Institute of Radiophysics and Electronics, 0203 Ashtarak, Armenia}
\affiliation{Center of Strong Fields Physics, Yerevan State University, 0025 Yerevan, Armenia}

\author{M.~E.~Veysman}
\affiliation{Joint Institute for High Temperatures RAS, Moscow, 125412, Russia}

\author{N.~E.~Andreev}
\affiliation{Joint Institute for High Temperatures RAS, Moscow, 125412, Russia}
\affiliation{Moscow Institute of Physics and Technology (State University), Moscow, 113303, Russia}

\author{H.~H.~Matevosyan}
\affiliation{Theoretical Physics Division, Institute of Radiophysics and Electronics, 0203 Ashtarak, Armenia}

\date{\today }

\begin{abstract}
A simple model for the dielectric function of a completely ionized plasma with an arbitrary
ionic charge, that is valid for long-wavelength high-frequency
perturbations is derived using an approximate solution of a linearized Fokker-Planck kinetic equation for  electrons with a Landau collision integral.
The model accounts for both the electron-ion collisions and
the collisions of the subthermal (cold) electrons with thermal ones. The relative contribution of the
latter collisions to the dielectric function is treated phenomenologically, introducing some parameter
$\varkappa$ that is chosen in such a way as to get a well-known expression for stationary electric conductivity
in the low-frequency region and fulfill the requirement of a vanishing contribution of electron-electron collisions
in the high-frequency region. This procedure ensures the applicability of our model in a wide range of plasma
parameters as well as the frequency of the electromagnetic radiation. Unlike the interpolation formula proposed
earlier by Brantov \emph{et al.}~[Brantov \emph{et al.}, JETP \textbf{106}, 983 (2008)], our model fulfills the
Kramers-Kronig relations and permits a generalization for the cases of degenerate and strongly coupled plasmas.
With this in mind, a generalization of the well-known Lee-More model~[Y.~T.~Lee and R.~M.~More, Phys. Fluids
\textbf{27}, 1273 (1984)] for stationary conductivity and its extension to dynamical conductivity~[O.~F.~Kostenko
and N.~E.~Andreev, GSI Annual Report No.~GSI-2008-2, 2008 (unpublished), p.~44] is proposed for the case of plasmas with arbitrary ionic charge.
\end{abstract}


\maketitle

\section{Introduction}
\label{sec:1}

The problem of interaction of intense laser pulses with solids and plasmas continues to be the
subject of intense experimental and theoretical research. These interactions are associated with
both the fundamental aspects of the behavior of matter in ultrastrong laser fields and various
applications such as fast ignition~\cite{Kitagawa2004IEEE}, the development of new sources of x-ray radiation
and warm dense matter production~\cite{Zastrau2010PRE}, particle acceleration~\cite{Carroll2010NJP},
and the laser generation of shock waves. In most part of these studies the high-power laser pulse
ionizes the matter so one eventually has to deal with a partially or fully ionized plasma.
In the past few decades much effort has been devoted to investigate the various aspects of laser-plasma
interactions (see, e.g., Refs.~\cite{Silin1973,SilinRukhadze13,Silin13,Alexandrov84}).
Currently various models of these interactions are widely discussed (see, e.g.,
Refs.~\cite{Pukhov2003RPP,andreevTVT03,VeysmanPoP06,VeysmanJPB08,Povarnitsyn12AppSS,Povarnitsyn12PoP} and references
therein). The key quantity which characterizes laser-matter interaction as well as the optical properties
of the matter is the plasma dielectric function (permittivity) $\varepsilon$, which determines the
electrodynamic response of the system on perturbations. Thus, the construction of the theoretical
models for the plasma permittivity valid in a wide range of the plasma parameters is of fundamental
and practical importance.

Plasma permittivity has been studied in detail and is well known in two limiting cases corresponding to the collisionless case based on the solution of the Vlasov kinetic equation \cite{Ichimaru73,Alexandrov84,SilinRukhadze13,Silin13} and to the strongly collisional hydrodynamic limit \cite{Braginskii65,Shkarofsky66}. In the latter regime the ranges of applicability of the corresponding expressions for the permittivity of a collisional plasma are strongly restricted and cannot be used for arbitrary values of $\omega /\nu_{e}$ and $k\lambda_{ei}$, where $\nu_{e}$ is the electron-ion collision frequency and $\lambda_{ei}$ is the mean free path of electrons with respect to their collisions with ions. An important development in recent years is the weakly collisional theory proposed in Ref.~\cite{Silin02UFN}, which extends the range of the analytical description of the permittivity for a collisional plasma compared to the collisionless case.

To obtain qualitative descriptions of collisional regimes of a plasma the Bhatnagar-Gross-Krook (BGK) \cite{Bhatnagar54} collisional model in the kinetic equation for electrons has been widely used with or without number-conservation procedure \cite{Alexandrov84,Mermin1970PRB,Nersisyan2004PRE,Nersisyan2008rev,LL_PhysKin,Clemmow90,Opher02PRE}. The appeal of this model is its simplicity, which in its original nonconserving form amounts to the replacement of $\omega\to \omega +i\nu$ in the argument of the plasma dispersion function, where $\nu$ is a model collision frequency. Furthermore, more advanced number- and energy-conserving BGK
as well as number-, momentum-, energy-conserving BGK
models have been presented in Refs.~\cite{Fried66,Selchow00NIMA} and \cite{Morawetz2000PRE,Atwal2002PRB}, respectively, which yield analytic expressions for the permittivities in terms of combinations of the plasma dispersion function.
However, for a completely ionized plasma, the model permittivity within the BGK
approximation and the corresponding Drude model for the transverse permittivity \cite{Alexandrov84,LL_PhysKin,Clemmow90,Opher02PRE}
lead to the significant deviations from the known limiting cases in the range of moderate and strong
collisions \cite{Brantov06IEEE,Bychenkov97PoP,Bychenkov98PPRep}. For instance, it has been found that this model cannot reproduce
the plasma permittivity in the strongly collisional hydrodynamic regime considered in Ref.~\cite{Shkarofsky66}. A significant improvement of the theory has been achieved within the Lorentz plasma model \cite{Bychenkov98PPRep,Koch75PhFluids,Penano97PoP}.
However, Lorentz plasma  model cannot describe permittivity accurately in a wide range of parameters even for a highly-ionized plasma, as long as the electron-electron collisions are neglected in this model. We also mention the model of Ref.~\cite{Selchow1999PRE} with a simplified
Fokker-Planck kinetic equation, where the diffusion tensor and the friction coefficient are treated as given constants. The resulting dielectric
function has been compared with the number-conserving Mermin dielectric function demonstrating that both functions are almost identical.

For the case of a plasma with a large ionic charge $Z\gg 1$, where the electron-electron collision integral is involved only
in the equation for the isotropic part of the electron distribution function, the longitudinal and
transverse permittivities have been obtained in Refs.~\cite{Brantov04PRL,Brantov05JETP} and \cite{Bychenkov97PoP}, respectively.
Generalization of the latter results to the case of an arbitrary ionic charge $Z$ requires, in addition, the consideration
of the electron-electron collision integral for the anisotropic part of the perturbed distribution function.
This problem has been considered recently in Ref.~\cite{Brantov08JETP} without any constraints on the parameters
under consideration. The model developed in Ref.~\cite{Brantov08JETP} is based on the solution of a linearized kinetic equation for
electrons with a Landau collision integral. In addition, the suggested method of solving the kinetic
equation is valid for an arbitrary ionic charge $Z$, an arbitrary relation between the perturbation
inhomogeneity scale length $k^{-1}$ and the electron mean free path, and an arbitrary relation between
the characteristic time scale $\omega^{-1}$, electron collision time, and the time scale of collisionless
electron motion $1/kv_{\mathrm{th}}$, where $v_{\mathrm{th}}$ is the thermal electron velocity.

However, the model proposed in Ref.~\cite{Brantov08JETP} being accurate in a wide range of parameters is rather complicated
and does not determine the permittivity in an explicit form expressed through the plasma parameters. Therefore, simplified
but still accurate models for the plasma permittivity are highly desirable. Besides, the model of Ref.~\cite{Brantov08JETP}
considers the case of ideal nondegenerate plasmas only, which restricts its use for description of laser-matter interaction
in a wide range of parameters.

In the present study we propose an alternative and simplified solution of the kinetic equation for electrons with a
Landau collision integral for an arbitrary charge of plasma ions. The model accounts for both the electron-ion collisions
and the collisions of the subthermal (cold) electrons with thermal ones. As has been shown in Ref.~\cite{Silin02UFN} the
latter collisions may considerably contribute in the common integral of collisions and one can derive an algebraic
expression for the respective part of the integral of electron-electron collisions containing, however, some free
parameter. This parameter is then adjusted so that to ensure the agreement of the present model with respective
expression for a stationary electric conductivity at low-frequencies~\cite{Balescu60PhFluids,Brantov08JETP} and proper
behavior of high-frequency conductivity (or permittivity) at high-frequencies.
Moreover, the presented model permits simple extensions for the cases of degenerate and/or strongly coupled plasmas,
which makes it possible to use it for description of optical properties of plasmas in a wide range of temperatures
and densities. Thus, this model represents the generalization of the well-known Lee-More model~\cite{Lee84PhysFluids}
for a stationary conductivity and its extension for a dynamical conductivity~\cite{Kostenko08GSI} (in the same
relaxation-time approximation). It is valid for plasmas with arbitrary degeneracy and arbitrary ionic charge, where
the electron-electron collisions play an essential role.

\section{Theoretical model}
\label{sec:2}

Within linear response approximation the evolution of the small perturbations arising in a homogeneous, collisional,
and unmagnetized plasma is considered below. The case of the long wavelength and high-frequency perturbations is
considered for electron component of plasma. The dynamics of the plasma ions is neglected. More specifically, we
assume that $kv_{\mathrm{th}}\ll \omega$, $k\lambda_{ei}\ll 1$ and $k\lambda_{ee}\ll 1$, where $k^{-1}$ is the
wavelength of the perturbations, $\omega^{-1}$ is the characteristic time, and $\lambda_{ei}$ ($\lambda_{ee}$) is the
mean free path of the electrons with respect to their collisions with ions (electrons).

The evolution of the electron component of the plasma is governed by the Fokker-Planck
kinetic equation for the velocity distribution function $f(\mathbf{v},t)$ of
the electrons. The distribution function of the ions is fixed and is
given by $f_{i}(\mathbf{v},t)=\delta (\mathbf{v}) $. Neglecting the spatial
inhomogeneity of the electron distribution function in the
case of the long wavelength perturbations, the kinetic equation can be written as \cite{SilinRukhadze13,Silin13,Ichimaru73}
\begin{equation}
\frac{\partial f}{\partial t}-\frac{e}{m}\mathbf{E}\cdot \frac{\partial f}{%
\partial \mathbf{v}}=J[f]\equiv \frac{\partial }{\partial v_{i}}\left( D_{ij}%
\frac{\partial f}{\partial v_{j}}-F_{i}f\right) ,
\label{f}
\end{equation}
where $J[f]=J_{ee}[f]+J_{ei}[f]$ is the collision term with the
contributions of the electron-electron $J_{ee}[f]$ and electron-ion
$J_{ei}[f]$ collisions, respectively, $\mathbf{E}$\ is the
self-consistent electric field strength, $D_{ij}$ and $\mathbf{F}$ are the diffusion
tensor and the friction force in a velocity space, respectively.

Taking the collision term $J[f]$ in the form of Landau
\cite{SilinRukhadze13,Silin13,Ichimaru73}, the velocity diffusion tensor and the
friction force are given by
\begin{eqnarray}
&&D_{ij} =\frac{h}{2}\left[ \frac{1}{Z}\int f(\mathbf{v}^{\prime },t)g_{ij}(%
\mathbf{u}) d\mathbf{v}^{\prime }+g_{ij}(\mathbf{v})\right] ,  \label{Dij}
\\
&&F_{i} =\frac{h}{2}\left[ \frac{1}{Z}\int f(\mathbf{v}^{\prime },t)
\frac{\partial g_{ij}(\mathbf{u})}{\partial u_{j}}d\mathbf{v}^{\prime }+
\frac{m}{m_{i}}\frac{\partial g_{ij}(\mathbf{v})}{\partial v_{j}}\right],
\label{Fi}
\end{eqnarray}
where $\mathbf{u}=\mathbf{v}-\mathbf{v}^{\prime }$,
\begin{equation}
g_{ij}(\mathbf{v})=\frac{1}{v}\left( \delta _{ij}-\frac{v_{i}v_{j}}{v^{2}}\right) ,
\label{gij}
\end{equation}
$\partial g_{ij}(\mathbf{v})/\partial v_{j}=-2v_{i}/v^{3}$, $\delta_{ij}$ is the unit tensor of rank 3,
$h=3\sqrt{\pi/2}\nu_e v_{\mathrm{th}}^3$,
\begin{equation}
\nu _{e}=\frac{4\sqrt{2\pi }n_{e}Ze^{4}}{3(mT^{3})^{1/2}}\Lambda
\label{nue}
\end{equation}
is the effective electron-ion collision frequency, and $v_{\mathrm{th}}=\sqrt{T/m}$.
Here $-e$, $m$, $n_{e}$ and $Ze$, $m_{i}$, $n_{i}$ are the electron and ion charges, masses and equilibrium densities,
respectively, $T$ is the temperature of electron component and $\Lambda $ is the Coulomb logarithm, which is defined later. Charge neutrality of the plasma with $n_{e}=Zn_{i}$ and an arbitrary (and finite) ionic charge $Z$ are assumed.

The first and the second terms in Eqs.~(\ref{Dij}) and (\ref{Fi}) correspond to the electron-electron
and electron-ion collisions, respectively. The last term in Eq.~(\ref{Fi})
describes the energy exchange between electrons and ions and is proportional
to the small parameter $\sim m/m_{i}\ll 1$. This term will be
neglected in the subsequent calculations.
The electron-electron collisions terms in Eqs.~(\ref{Dij}) and (\ref{Fi}) contain the
inverse $Z^{-1}$ of the ionic charge number $Z$. Hence, these terms vanish
at the limit $Z\gg 1$ of the highly ionized ions and one arrives at the Lorentz plasma model \cite{LL_PhysKin} in this case, which is frequently used in hydrodynamic codes due to its simplicity~\cite{andreevTVT03,VeysmanPoP06,VeysmanJPB08,Povarnitsyn12AppSS,Povarnitsyn12PoP}.

Lorentz model is justified only for plasma with highly ionized ions with $Z\gtrsim 10$.
For plasmas with $Z<10$ electron-electron collisions should be accounted for numerically more precise calculations: though due to the momentum conservation
(i.e. $\int \mathbf{v}J_{ee}[f]d\mathbf{v}=0$) they do not directly contribute to the induced current density. Nevertheless, they modify electron distribution function and thus influence on the value of permittivity.
Rigorous kinetic theory for calculation of permittivity of plasma with account for electron-electron
collisions and nonlocal transport was proposed in Ref.~\cite{Brantov08JETP}.

In the present paper more simple, but physically motivated approach is considered, which makes one possible to derive simple expression for permittivity of plasmas with account for contribution of electron-electron collisions and permits further generalizations for quantum plasmas and/or for strongly coupled plasmas. Unlike interpolation formula proposed in Ref.~\cite{Brantov08JETP}, present model fulfills Kramers-Kronig relations and permits further extension for degenerate plasma case.

In order to derive this model let us note, that in accordance with Ref.~\cite{Silin02UFN},
the effective frequency for collisions of the subthermal (cold) electrons (with velocities $v\ll v_{\mathrm{th}}$) with the thermal ones (with $v\sim v_{\mathrm{th}} $) behaves as $\nu_{c,ee}\sim (v_{\mathrm{th}}/v)^{3}\gg \nu_{ee}$ so it considerably exceeds the similar frequency $\nu_{ee}$ for the collisions of the thermal electrons. Therefore, even in a weakly collisional plasma the cold electrons experience strong collisions
with the thermal ones and may essentially contribute to the coefficients~(\ref{Dij}) and (\ref{Fi}). Taking in mind this, we restrict the upper limits of the velocity
integrations in Eqs.~(\ref{Dij}) and (\ref{Fi}) by some value $v_{m}\lesssim v_{\mathrm{th}}$. Also since $v\simeq v_{\mathrm{th}}$ in Eqs.~(\ref{Dij}) and (\ref{Fi}), the tensor $g_{ij}(\mathbf{u})$ and the vector $\partial g_{ij}(\mathbf{u})/\partial u_{j}$ can be replaced by $g_{ij}(\mathbf{v})$ and $\partial
g_{ij}(\mathbf{v})/\partial v_{j}$, respectively, taking them out from the
$\mathbf{v}^{\prime }$-integrals in Eqs.~(\ref{Dij}) and (\ref{Fi}).

Next, within linear response approach the distribution function
$f(\mathbf{v}^{\prime },t)$ in Eqs.~(\ref{Dij}) and
(\ref{Fi}) can be replaced by the equilibrium distribution function of the
electrons $f_{0}(v^{\prime })$, and taking in mind affirmations stated above,
 $f_{0}(v^{\prime })$ can be replaced by $f_{0}(v^{\prime })\simeq
f_{0}(0)$. As a result from Eqs.~(\ref{Dij}) and (\ref{Fi}) we obtain
\begin{eqnarray}
&&D_{ij} =\frac{h}{2}\left( 1+\frac{1}{Z_{\ast }}\right) g_{ij}(\mathbf{v}), \label{Dij2} \\
&&F_{i} =\frac{h}{2Z_{\ast }}\frac{\partial g_{ij}(\mathbf{v})}{\partial v_{j}},  \label{Fi2}
\end{eqnarray}
where $Z_{\ast }=Z/\varkappa $ with $\varkappa =\frac{4\pi }{3}v_{m}^{3}f_{0}(0)$.

It is seen that the contribution of the electron-electron collisions (the terms containing
the effective charge number $Z_{\ast }$) is not negligible in
the coefficients~(\ref{Dij2}) and (\ref{Fi2}). The parameter $\varkappa $
introduced above is the relative fraction of the slow electrons contributing to
the coefficients (\ref{Dij2}) and (\ref{Fi2}). Clearly $\varkappa
\lesssim 1$ which results in $Z_{\ast }>Z$, i.e. a larger effective charge of
the ions compared to $Z$.

To obtain an equation for perturbed distribution function one can substitute
$f=f_{0}+f_{1}$ (with $f_{1}\ll f_{0}$) into~(\ref{f}) to get the equation
\begin{equation}
-i\omega f_{1\omega }(\mathbf{v})-\frac{e}{m}(\mathbf{E}_{\omega }\cdot
\mathbf{v})\frac{1}{v}f_{0}^{\prime }(v)=J\left[ f_{1\omega }(\mathbf{v})\right] ,
\label{fomega}
\end{equation}
for Fourier transform with respect to the time $t$ of the perturbed distribution function $f_{1}$. Here
$\mathbf{E}_{\omega }$ is the Fourier transform of electric field; prime indicates the
derivative with respect to the argument.
The equilibrium distribution function in unperturbed state is assumed to be
isotropic $f_{0}=f_{0}(v)$.

In order to solve Eq.~(\ref{fomega}) it is convenient to introduce a new unknown
and isotropic function $\Phi_{\omega }(v)$ via the relation
\begin{equation}
f_{1\omega }(\mathbf{v})=\frac{e}{m\omega }(\mathbf{E}_{\omega }\cdot \mathbf{v})\Phi_{\omega }(v).
\label{fPhi}
\end{equation}
This relation~(\ref{fPhi}) explicitly separates the isotropic [the term
$\Phi _{\omega }(v)$] and anisotropic [the term $(\mathbf{E}_{\omega }\cdot
\mathbf{v})$] parts of the distribution function $f_{1\omega }(\mathbf{v})$.
Note that such a choice for the perturbed distribution function is
stimulated by the structure of~(\ref{fomega}). Then inserting
equation~(\ref{fPhi}) into~(\ref{fomega}) and using the diffusion tensor (\ref{Dij2})
and the friction force (\ref{Fi2}) yields after straightforward
calculations an ordinary differential equation for the unknown function $\Phi _{\omega }(v)$
\begin{equation}
\frac{1}{\omega Z_{\ast }}\Phi _{\omega }^{\prime }(v)+\frac{i }{hv}\left(
v^{3}+ih/\omega \right) \Phi _{\omega }(v)=-\frac{1}{h}vf_{0}^{\prime }(v).
\label{eq:Phi}
\end{equation}

An expression similar to Eq.~(\ref{eq:Phi}) has been considered previously
in Refs.~\cite{Bychenkov97PoP,andreevTVT03,VeysmanPoP06,VeysmanJPB08,Povarnitsyn12AppSS,Povarnitsyn12PoP} neglecting, however, the
first term containing the derivative of the function $\Phi _{\omega }(v)$, that is
justified for $Z\gg 1$.
In this case the differential equation (\ref{eq:Phi}) is reduced to an
algebraic one with a simple solution
\begin{equation}
\Phi _{\omega }^{(\mathrm{L})}\left( v\right) =i\frac{v^{2}f_{0}^{\prime }
\left( v\right) }{v^{3}+ih/\omega }
\label{PhiZinf}
\end{equation}
which eventually yields the Lorentz model for plasma permittivity
\cite{LL_PhysKin,Bychenkov97PoP,andreevTVT03,VeysmanPoP06,VeysmanJPB08,Povarnitsyn12AppSS,Povarnitsyn12PoP}.
For an arbitrary charge state $Z$ of the plasma ions and for a finite
parameter $\varkappa $, the solution of Eq.~(\ref{eq:Phi}) is given by
\begin{equation}
\Phi _{\omega }(v)=\frac{Z_{\ast }\omega}{h}\int_{v}^{\infty }\exp \left[ \frac{%
iZ_{\ast }\omega }{3h}\left( u^{3}-v^{3}\right) \right] \left( \frac{v}{u}%
\right) ^{Z_{\ast }}f_{0}^{\prime }(u)udu.
\label{Phisol}
\end{equation}
The perturbations of the current induced in the plasma by the electric field
$\mathbf{E}$ are determined by $\mathbf{j}_{1}=-n_{e}e\int \mathbf{v}f_{1}(%
\mathbf{v},t)d\mathbf{v}$. The Fourier transform of this quantity is then
given by
\begin{equation}
\mathbf{j}_{1\omega }=-\frac{\omega _{p}^{2}}{4\pi \omega}\int \mathbf{v}(\mathbf{E}_{\omega }\cdot
\mathbf{v})\Phi _{\omega }(v)d\mathbf{v} ,
\label{jomega}
\end{equation}
where $\omega _{p}^{2}=4\pi n_{e}e^{2}/m$ is the plasma frequency. Using
this relation one can calculate the conductivity tensor and hence
the permittivity tensor of the collisional electron plasma which can be
represented in the form $\varepsilon _{ij}(\omega )=\varepsilon (\omega
)\delta _{ij}$ with
\begin{eqnarray}
&&\varepsilon (\omega )=1-\frac{\omega _{p}^{2}}{\omega^2 }K_{0} (\omega ) , \nonumber \\
&&K_0(\omega )=\frac{4\pi i}{3} \int_{0}^{\infty }\Phi _{\omega }(v)v^{4}dv . \label{eps}
\end{eqnarray}
The obtained expression together with the distribution function (\ref{Phisol})
determines the high-frequency dielectric function of the collisional plasma
for an arbitrary effective charge $Z_{\ast }$ of the ions.
The expression\rf{eps} can be further simplified if Eq.~(\ref{Phisol})
is inserted into it and one performs an integration by parts. This yields
\begin{equation}
K_0 (\omega )=\dst\frac{i \chii}{\xi_\omega} \frac{8\sqrt{2}\pi}{3}v_{\mathrm{th}}^3
\int_{0}^{\infty } \! F\left(1;\alphaz ;i\betaz \xi^3 \right)\xi^6  f_0'(\xi) d\xi,
\label{eps2}
\end{equation}
where $f'_{0}(\xi)$ denotes derivative of $f_{0} (\xi)$ over $\xi$,
\begin{equation}
\xi = \dst \frac{v}{\sqrt{2}v_{\mathrm{th}}}, \,\,
\xi_{\omega}=\frac{3\sqrt{\pi }}{4}\frac{\nu _{e}}{\omega}, \,\,
\alphaz = \frac{\Zef+8}{3}, \,\,
\betaz = \frac{\Zef }{3 \xi_\omega} ,
\label{eq:par}
\end{equation}
and $F(a;b;z)$ is the confluent hypergeometric function.
Using the properties of the confluent hypergeometric functions (see, e.g., Ref.~\cite{Gradshteyn80}) one can write
the series expansion for $F$ over its third argument for the case $\betaz \xi^3 \ll 1 $,
\begin{equation}
\dst F\left(1;\alphaz ;i\betaz \xi^3\right) = 1 + i\frac{\betaz\xi^3}{\alphaz} -
\frac{\betaz^2\xi^6}{\alphaz(\alphaz+1)}+\ldots,
\label{Fw0}
\end{equation}
and the asymptotic expression for $F$ over the value of $\Zef^{-1}$,
 $\Zef \gg 1 $:
\begin{equation}
F\left(1;\alphaz ;i\betaz \xi^3\right) = \dst\frac{1}{1-\betazz} + \sum\limits_{n\geqslant  1}
\frac{1}{\Zef^n}\frac{\betazz P_n(\betazz)}{(1-\betazz)^{2n+1}} ,
\label{FZinf}
\end{equation}
where $\betazz = i\xi^3/\xi_\omega$ and $P_n(\betazz)$ are polynomials of $\betazz$ of the power $n$.
The first three have the following values:
\begin{eqnarray}
&&P_{1} = 5\betazz-8, \\[2ex]
&&P_{2} = 10\betazz^{2}-47\betazz +64, \\[2ex]
&&P_{3} =-10\betazz^{3}+48\betazz^{2}+69\betazz-512.
\label{R}
\end{eqnarray}
Considering Eq.\rf{Fw0}, one can derive from Eq.\rf{eps2} the following expression for the function $K_{0}(\omega)$ in the limiting case of low frequencies $\omega \ll \nu_e$:
\begin{equation}
K_0(\omega )=\frac{3\chir}{\xi_{\omega}^{2}}\langle \xi^{6}\rangle
-\frac{2i\chii}{\xi_{\omega}}  \langle \xi^{3}\rangle ,
\label{eps_w0}
\end{equation}
where $\langle \xi^n \rangle $ indicates an average of the value $\xi^n $ over the unperturbed distribution function $f_{0}(\xi)$ and two parameters $\chir$ and $\chii$ depend on the effective charge $Z_{\ast }$ as follows:
\begin{equation}
\chir= \frac{1}{(1+5/\Zef)(1+8/\Zef)},\quad \chii=\frac{1}{1+5/\Zef}.
\label{chi}
\end{equation}
Considering Eq.\rf{FZinf}, one can derive from Eq.\rf{eps2} the expression for the function $K_{0}(\omega)$ in the opposite limiting case of high frequencies $\omega \gg \nu_e$:
\begin{equation}
K_0(\omega )= 1-  i\chiii   \frac{8\pi \sqrt{2}}{3} v_{\mathrm{th}}^3 \xi_\omega f_{0}(\xi=0) ,
\label{eps_winf}
\end{equation}
where the parameter
\begin{equation}
\chiii =1+2/Z_{\ast }
\label{chiii}
\end{equation}
contains dependence on the effective charge $Z_{\ast }$.
Equations~(\ref{eps_w0}) and (\ref{eps_winf}) represent well-known cases for the normal low-frequency and normal high-frequency skin effects, respectively.
It should be emphasized that they depend essentially on the ion effective
charge $Z_{\ast }$ and they are valid for arbitrary equilibrium distribution function $f_0$, including one for the degenerate electron plasma. Below these limiting cases will be used for determination of the unknown parameter $\varkappa = Z/\Zef$.

\subsection{Nondegenerate electron plasma}

For the Maxwell equilibrium distribution function $f_0(\xi) = (2\pi v^{2}_{\mathrm{th}})^{-3/2} e^{-\xi^2} $ one has from Eq.\rf{eps2}
the following expression:
\begin{equation}
K_0 (\omega )= \frac{}{} \frac{-8i\chii}{3\xi_{\omega} \sqrt{\pi}}
\int_{0}^{\infty } \!  F\left(1;\alphaz ;i\betaz \xi^3\right)
\xi^7 e^{-\xi^2}d\xi.
\label{eps_Maxw}
\end{equation}
The limiting cases~(\ref{eps_w0}) and (\ref{eps_winf}) for the case of the Maxwell distribution function give, respectively,
\begin{equation}
K_0(\omega )=\frac{315}{8} \frac{\chir}{\xi^{2}_{\omega}}-
\frac{8i}{ \sqrt{\pi}}\frac{\chii}{\xi_{\omega}}
\label{eps_w0_Maxw}
\end{equation}
 and
\begin{equation}
K_0(\omega )= 1-i\frac{4}{3\sqrt{\pi}} \xi_\omega \chiii ,
\label{eps_winf_Maxw}
\end{equation}
which completely agree with the standard forms of the corresponding
expressions~\cite{SilinRukhadze13,Ichimaru73,Silin13,andreevTVT03,VeysmanPoP06,VeysmanJPB08,Povarnitsyn12AppSS,Povarnitsyn12PoP} in the case $\chir=\chii=\chiii=1$, which follows from Eqs.\rf{chi} and\rf{chiii} in the formal limit $Z\rightarrow \infty $.
Inserting the first term of Eq.\rf{FZinf} into Eq.~(\ref{eps_Maxw}) one gets the Lorentz model for optical properties of plasmas:
\begin{equation}
K_0 (\omega )= \frac{8\chii}{3\sqrt{\pi}} \int_{0}^{\infty } \!
\frac{\xi^7 e^{-\xi^2}}{\xi^3 + i\xi_\omega}d\xi,
\label{eps_MaxwL}
\end{equation}
considered previously (for $\chii=1$)
in Refs.~\cite{Bychenkov97PoP,andreevTVT03,VeysmanPoP06,VeysmanJPB08,Povarnitsyn12AppSS,Povarnitsyn12PoP}.

In order to use Eq.\rf{eps2} or \rf{eps_Maxw}, one has to derive an expression for the relative fraction
$\varkappa$ of electron-electron collisions with subthermal electrons. This can be done if one takes into account the
above limiting cases.
(i) For $\omega \to \infty$ the permittivity does not depend on electron-electron collisions~\cite{Silin13,SilinRukhadze13,LL_PhysKin,Brantov08JETP},
which means that it should not contain a dependence on $\Zef $. Recalling Eqs.\rf{chiii} and\rf{eps_winf}, this means that
\begin{equation}
\Zef \to \infty \;\;\;\mbox{ for }\; \omega \to \infty .
\label{Zeffwinf}
\end{equation}
ii) For $\omega \to 0$ one has the respective interpolation formula for stationary conductivity
\begin{equation}
\sigma_0=\gamma_\sigma(Z)\sigma_{\mathrm{sh}}, \quad \sigma_{\mathrm{sh}}=\frac{2}{\pi^{3/2}}\frac{\omega_p^2}{\omega\xi_\omega},\quad \gamma_\sigma=\frac{a+Z}{b+Z},
\label{sigma0}
\end{equation}
where $a=0.87$ and $b=2.2$ (see Refs.~\cite{Balescu60PhFluids,Brantov08JETP}). Considering the connection
\begin{equation}
\sigma' = -\frac{\omega_p^2}{4\pi \omega} \mathrm{Im}[K_{0}(\omega)]
\label{sigma_K0}
\end{equation}
of the real part of conductivity and the function $K_{0} (\omega)$, one can write
 the following expression for the imaginary part of $K_0 (\omega)$ in the stationary case: $\mathrm{Im}[K_0 (\omega)] \vert_{\omega\to 0}
=-8i\gamma_\sigma/\sqrt{\pi}\xi_\omega$. Comparing this expression with Eqs.\rf{eps_w0_Maxw} and\rf{chi}, one gets
\begin{equation}
\varkappa(Z,\omega\to 0)=\frac{Z}{\Zef(\omega\to 0)} = \frac{Z(b-a)}{5(Z+a)}.
\label{Zeffw0}
\end{equation}
Taking into account Eqs.\rf{Zeffwinf} and\rf{Zeffw0}, one can propose the following interpolation for
$\varkappa (\omega)$ in the whole frequency range:
\begin{equation}
\varkappa (\omega) = \varkappa_0\left[1 + (C/\xi_\omega)^s\right]^{-1},
\label{kappa}
\end{equation}
where $\varkappa_0=\varkappa(\omega\to 0)$ is given by Eq.\rf{Zeffw0} and $C>0$ and $s>0$ are positive
numerical constants, which can be withdrawn, for example, from the comparison with the exact calculations.

\subsection{Degenerate electron plasma}

In this section we generalize the permittivity\rf{eps2} obtained for a nondegenerate electron
plasma to the cases of a partially or fully degenerate plasma. Strictly speaking the starting point in this case should be
the quantum kinetic equation. However, below arguments show that simple generalization of Eq.\rf{eps2} is possible in the manner analogous to that done for the case of Lorentz plasma with arbitrary degeneracy in Refs.~\cite{Lee84PhysFluids,Kostenko08GSI}.

First, it has been shown previously (see, e.g., Ref.~\cite{Reinholz12PRE}), that the calculation of velocity-dependent electron-ion
collision frequency $\nu (v)$ [$\nu (v)\sim h/v^{3}$, where $h$ has been introduced in Sec.~\ref{sec:2}] on the basis of
the quantum kinetic equation yields the same result, as if one starts from the classical kinetic equation, where, however,
the classical Coulomb logarithm has to be replaced by the quantum one.  Second,
the electron-electron collisions in a degenerate plasma have been investigated in detail in Refs.~\cite{Lampe68a,Lampe68b,Flowers76,Shternin06}
using quantum kinetic equation approach. However, starting from the quantum kinetic equation and following the same steps
that led to Eqs.~(\ref{Dij2}) and (\ref{Fi2}) we now get the similar expressions. Finally, it is well
known (see, e.g., Refs.~\cite{Alexandrov84,Silin13}) that at vanishing quantum recoil with $\hbar k^{2}/2m\ll \omega$, the
dielectric function which follows from the collisionless quantum kinetic equation in a random-phase approximation \cite{Lindhard54}
is identical to the corresponding classical expression. Thus, in the case of a degenerate plasma Eq.\rf{eps2} is applicable assuming that $\hbar k^{2}/2m\ll \omega$ in addition to the conditions introduced at the beginning of
Sec.~\ref{sec:2}.

In the case of a partially degenerate electron plasma the equilibrium distribution function $f_{0}(\xi )$ in Eq.\rf{eps2}
is given by the Fermi-Dirac distribution
\begin{equation}
f_{0}(\xi) = B_0 \fF(\xi), \,\,\,\, \fF(\xi) = \left[1+\exp(\xi^2-\emu)\right]^{-1} ,
\label{FFermy}
\end{equation}
where $B_0=(3/4\pi )(m/2 E_{\mathrm{F}})^{3/2}$ is the normalization constant, $E_{\mathrm{F}} =\frac{\hbar^{2}}{2m} (3\pi^{2}
n_{e})^{2/3}$ is the Fermi energy, $\emu = \mu/T$, $\mu$ is the chemical potential. Inserting the distribution~(\ref{FFermy})
into Eq.\rf{eps2} we arrive at
\begin{equation}
K_{0} (\omega ) =\dst\frac{-2i \chii}{\xi_{\omega} \varepsilon^{3/2}_{\mathrm{F}}}
\int_{0}^{\infty } \! F\left(1;\alphaz;i\betaz \xi^3 \right)
\fF(\xi)[1-\fF(\xi)] \xi^7 d\xi
\label{epsFermy}
\end{equation}
for a partially degenerate electron plasma with $\varepsilon_{\mathrm{F}} =E_{\mathrm{F}}/T$. It should be emphasized
that the definitions of the dimensionless quantities $\xi_{\omega}$ and $\betaz$ (see Eq.\rf{eq:par})
in Eq.~(\ref{epsFermy}) should contain now quantum expression for Coulomb logarithm $\Lambda$ in the expression for collision
frequency, Eq.\rf{nue}.

The dimensionless chemical potential in expression for $\fF$ is calculated from equation
\begin{equation}
\emu = X_{1/2} \left(\frac{2}{3} \varepsilon^{3/2}_{\mathrm{F}}\right),
\label{mu}
\end{equation}
where $X_{1/2}$ is the function inverse to the Fermi integral $F_{1/2}(x)$,
$X_{1/2}(F_{1/2}(x)) =x$, where $F_\alpha (x) = \int_0^\infty  t^\alpha (1+e^{t-x})^{-1} dt$.
For the numerical evaluation of Eq.~(\ref{mu}) it is useful to use the highly accurate
rational function approximations for the Fermi integrals and their inverse functions
derived in Ref.~\cite{Antia93}.

To compar the present approach with the previously known models it is also constructive to
consider some particular cases of the general expression~(\ref{epsFermy}). In the case of a
highly degenerate electron plasma with $T\ll E_{\mathrm{F}}$ the function~(\ref{epsFermy}) is
simplified and is given by
\begin{equation}
K_{0}(\omega ) =-\frac{i\chii}{\eta_{\omega}} F\left(1;\alpha_{Z};i\gamma_{Z}\right) .
\label{epsFermyD}
\end{equation}
Here $\gamma_{Z}=Z_{\ast}/3\eta_{\omega}$ and $\eta_{\omega }=\xi_\omega/\varepsilon^{3/2}_{\mathrm{F}} = \nu_d/\omega $, where
$\nu_d=(4Zme^{4}/3\pi \hbar^{3})\Lambda_d$ is the electron-ion
collision frequency in the case of a fully degenerate electron plasma derived by Flowers and Itoh~\cite{Flowers76}
and lately by Shternin and Yakovlev~\cite{Shternin06}, and  $\Lambda_d$ is the corresponding Coulomb logarithm.

Taking in mind, that for $E_F>T$ one has $\Zef\gg 1$ (see below), one can use expansion\rf{FZinf} for calculation of the confluent hypergeometric function in Eq.\rf{epsFermyD}. With only first term in this expansion one gets from Eq.\rf{epsFermyD}
\begin{equation}
K_{0}(\omega ) = \frac{1}{1+i\eta_{\omega}} ,
\label{epsFermyHD}
\end{equation}
i.e. the Drude expression for the function $K_{0}(\omega)$.

In the limit of low frequencies $\omega \ll \nu_e$ one can obtain from Eq.\rf{eps_w0} the expression for degenerate plasma similar
for that for nondegenerate one\rf{eps_w0_Maxw}:
\begin{equation}
K_0(\omega )= \frac{3\chir}{\xi^{2}_{\omega}}\frac{F_{7/2}(\emu)}{F_{1/2}(\emu)}  -
\frac{2i\chii}{\xi_{\omega}}\frac{F_{2}(\emu)}{F_{1/2}(\emu)},
\label{eps_w0_F}
\end{equation}
which in the limit $T\ll E_{\mathrm{F}}$ turns into
\begin{equation}
K_0(\omega )= \chir/\eta^{2}_{\omega} -i\chii/\eta_{\omega}.
\label{eps_w0_FD}
\end{equation}
Note that this result follows also from Eq.\rf{epsFermyD}.

From Eqs.\rf{eps_w0_FD} and\rf{sigma_K0}
one can obtain the following expression for the real part of stationary electric conductivity  $\sigma'(\omega\to 0)$ of highly-degenerate plasma (at $T \ll E_{\mathrm{F}}$):
\begin{equation}
\sigma' = \dst\frac{\chii }{\hbar} \frac{\sqrt{E_{\mathrm{F}}^3/E_{\mathrm{H}}}}{\sqrt{2}\pi Z}\frac{1}{\Lambda_d},
\label{sigmaD}
\end{equation}
where $E_{\mathrm{H}} = m e^4/\hbar^{2} \simeq 27.2$~eV is the Hartree energy.
This expression coincides with the generalization of the well-known Ziman formula~\cite{Ziman61Phmag}
for the partially degenerate case~\cite{Selchow00NIMA}, if one uses expression
\begin{equation}
\Lambda_d = \int_0^{\infty} \frac{S(k)}{k} \frac{f_{\mathcal{F}}(k\lambdabar ) dk}{|\varepsilon_{\mathrm{L}}(k,0)|^2}
\label{LZiman}
\end{equation}
for the Coulomb logarithm $\Lambda_d$ and put $\chii=1$ in Eq.~(\ref{sigmaD}). In Eq.\rf{LZiman} $\lambdabar =\hbar/(2mT)^{1/2}$
is the thermal wavelength, $S(k)$ is the static structure factor, and $\varepsilon_{\mathrm{L}}$ is the Lindhard dielectric
function~\cite{Lindhard54} for partially degenerate electron gas~\cite{Gouedard1978JMP,Arista84PhysRevA}.
In the opposite limiting case of high-frequencies, $\omega \gg \nu_e$, from Eq.\rf{eps_winf} one can obtain the expression
\begin{equation}
K_0(\omega )= 1 - i\chiii  \xi_{\omega} \varepsilon_{\mathrm{F}}^{-3/2} \left(1+e^{-\emu}\right)^{-1},
\label{eps_winf_F}
\end{equation}
which in the case of high degeneracy with $E_{\mathrm{F}} \gg T$ becomes
\begin{equation}
K_0(\omega )= 1 - i\chiii  \eta_{\omega} .
\label{eps_winf_D}
\end{equation}
Next, in the limit $Z_{\ast} \gg 1$, taking the first term of Eq.\rf{FZinf}, in the leading order one gets
from Eq.\rf{epsFermy} the following expression:
\begin{equation}
K_0 (\omega ) =\dst\frac{2\chii}{\varepsilon^{3/2}_{\mathrm{F}}}
\int_{0}^{\infty } \frac{\fF(\xi)[1-\fF(\xi)]}{\xi^3+i\xi_{\omega}} \xi^{7} d\xi,
\label{epsFermyL}
\end{equation}
which in the particular case $\chii=1$ coincides with a result, obtained in Refs.~\cite{Lee84PhysFluids,Kostenko08GSI} for the
electron conductivity of Lorentz plasma.

As mentioned above for accurate numerical treatment of the permittivity of degenerate plasmas one should use a proper expression for the Coulomb logarithm in Eq.\rf{nue} (and hence in Eqs.\rf{eq:par} and\rf{epsFermy}). For moderate values of degeneracy parameter $\Theta = \varepsilon^{-1}_{\mathrm{F}} = T/E_{\mathrm{F}} \gtrsim 1$ the wide-range formula for stationary electric conductivity for hydrogen-like plasmas ($Z=1$) was proposed in Ref.~\cite{Esser2003CPP}. Comparing the expression for $\sigma'$ obtained in Ref.~\cite{Esser2003CPP} and Eq.\rf{sigma0} for $Z=1$ and for weakly-degenerate plasma ($\Theta \gg 1$), one can use the following interpolation expression for $\Lambda$ in a wide range of density and temperature:
\begin{equation}
\Lambda (\Gamma, \Theta)= \frac{1/2}{1+{b_1/\Theta^{3/2}}}
\left[D\ln(1+A+B)-C- \frac{b_2}{b_2+\Gamma\Theta} \right] ,
\label{LERR}
\end{equation}
where $\Gamma = (4\pi n_{e}/3)^{1/3}Ze^2/T $ is the coupling parameter. The quantities $A$, $B$, $C$ and $D$
are functions of the parameters $\Gamma$ and $\Theta$ and are given by
\begin{eqnarray}
&&A = \frac{\Gamma^{-3} \left[1+a_4/(\Gamma^2 \Theta)\right]}{1 +a_2/(\Gamma^2 \Theta) +a_3/(\Gamma^2 \Theta)^2} \;
\left[a_1+c_1 \ln (c_2 \Gamma^{3/2}+1) \right]^2
\,, \quad
\nonumber\\[2ex]
&&B = \frac{b_3(1+c_3 \Theta)}{\Gamma \Theta(1+c_3 \Theta^{4/5})}
\,,  \qquad
C = \frac{c_4}{\ln(1+\Gamma^{-1})+c_5 \Gamma^2 \Theta} \,,
\nonumber\\[2ex]
&&D = \frac{\Gamma^{3}+a_5(1+a_6 \Gamma^{3/2})}{\Gamma^{3}+a_5}
 \,,\nonumber
\end{eqnarray}
with a set of numerical constants $a_0=0.03064$, $a_1=1.1590$,
$a_2=0.698$, $a_3=0.4876$, $a_4=0.1748$, $a_5=0.1$, $a_6=0.258$,
$b_1=1.95$, $b_2=2.88$, $b_3=3.6$, $c_1=1.5$, $c_2=6.2$, $c_3=0.3$,
$c_4=0.35$, $c_5=0.1$ (see Ref.~\cite{Esser2003CPP} for details).
The expression\rf{epsFermy} with Coulomb logarithm given by Eq.\rf{LERR} gives accurate description of permittivity of
plasmas for $Z=1$ and for $Z\gg 1$, where it goes into the Lorentz model of Lee and More~\cite{Lee84PhysFluids} for
stationary conductivity and its extension for dynamical conductivity~\cite{Kostenko08GSI}.

For highly and moderately degenerate plasmas the influence of electron-electron collisions will be decreased due to Pauli blocking~\cite{Esser2003CPP}. This effect can be taken into account, if one uses the expression for Spitzer factor in a degenerate electron plasma~\cite{Adams2007PhPl,Stygar2002PRE}:
\begin{equation}
\widetilde{\gamma}_\sigma (Z) =
\gamma_\sigma (Z) + \dst\frac{1-\gamma_\sigma (Z)}{1+0.6\ln\left(1+\Theta/20\right)}
\label{sigma0D}
\end{equation}
instead of respective expression for nondegenerate Spitzer factor $\gamma_\sigma (Z)$, Eq.\rf{sigma0}. In Ref.~\cite{Adams2007PhPl}
it was demonstrated, that the interpolation formula\rf{sigma0D} gives results very similar to those obtained by rigorous
quantum statistical approach.

Using the same arguments, which were used for derivation of expression\rf{Zeffw0}, one can obtain the following expression for the value of $\varkappa_0=Z/\Zef(\omega\to 0)$ for the case of partially or fully degenerate plasmas:
\begin{equation}
\varkappa_0=  Z \left[\widetilde{\gamma}_\sigma^{-1}(Z)-1\right]/5,
\label{Zeffw0D}
\end{equation}
where $\widetilde{\gamma}_\sigma$ is given by Eq.\rf{sigma0D}. The frequency dependence of $\varkappa$ is given by the same Eq.\rf{kappa}, as in the case of degenerate plasma.

It should be also mentioned, that the theoretical model described above is valid for frequencies
$\omega \lesssim \omega_p$. For frequencies higher than the plasma frequency the value of the real part of the function $K_{0}(\omega )$ will be considerably decreased, in comparison with one for $\omega < \omega_p$~\cite{DawsonOberman62PhFluids,DeckerMori94PhPl,Reinholz00PRE} as long as a charged particle screening at plasma frequency is replaced by the screening at laser frequency for $\omega > \omega_p$. This can be approximately accounted for by replacing $\omega_p$ by $\omega$ in Coulomb logarithm for the case $\omega > \omega_p$~\cite{DeckerMori94PhPl}.

\begin{figure}[tbp]
\includegraphics[width=80mm]{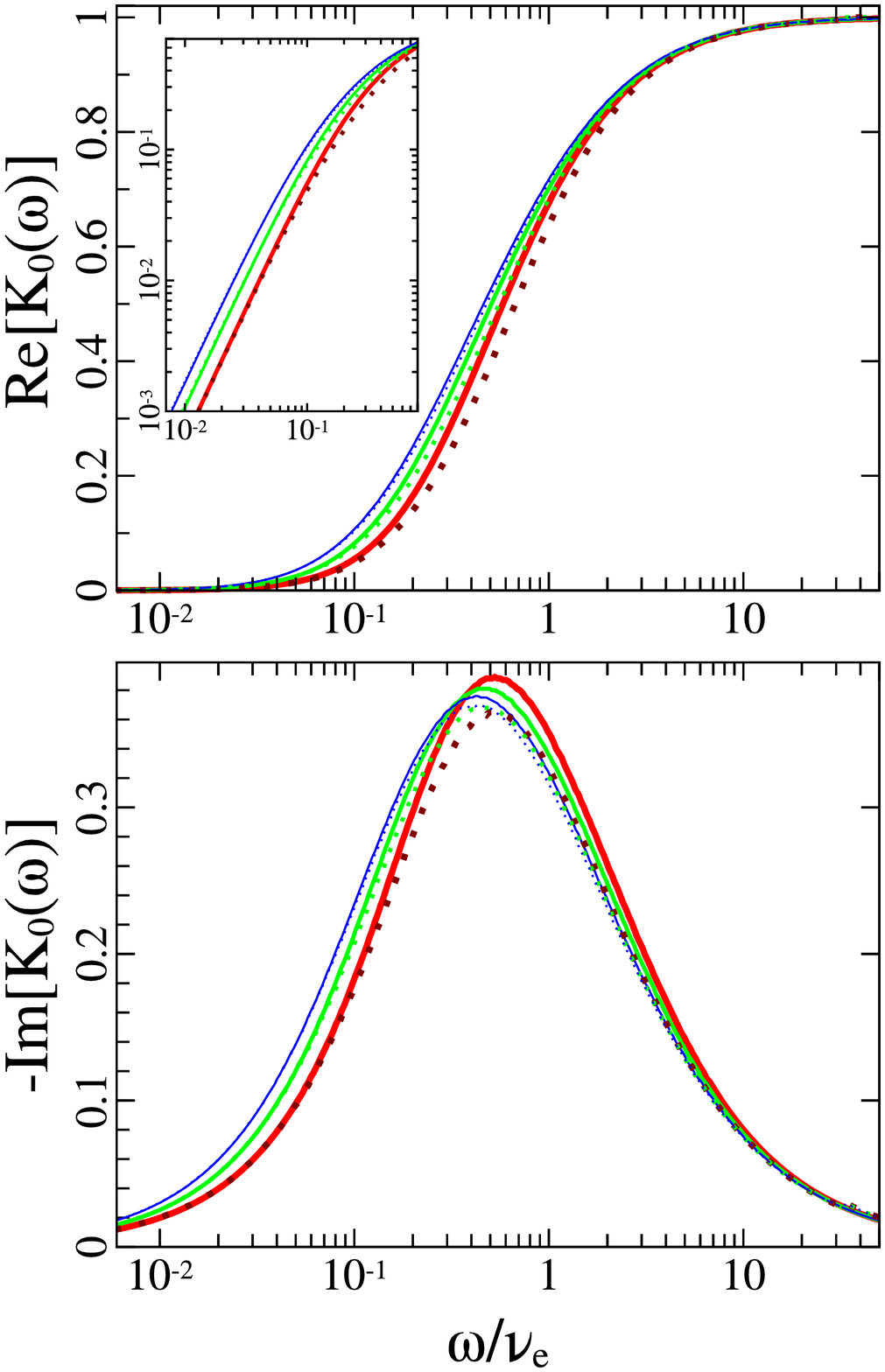}
\caption{(Color online) Real (top) and imaginary (with minus sign) (bottom) parts of $K_{0}(\omega )$ for
the nondegenerate electron plasma with different ionic charges $Z=1$ (thick lines), $Z=3$ (thinner lines),
and $Z=10$ (thinnest lines), calculated by Eqs.\rf{eps_Maxw} and\rf{kappa} with $C=s=1$ (solid lines) and by interpolation formula of Brantov \emph{et. al.}~\cite{Brantov08JETP} (dotted lines).}
\label{fig:K0_w_nu}
\end{figure}

\section{Numerical results}
\label{sec:3}

In\f{K0_w_nu} the results of the numerical calculations of the real $\mathrm{Re} [K_0(\omega)]$
and imaginary (with a minus sign) $-\mathrm{Im} [K_0(\omega)]$ parts of the function $K_{0}(\omega )$
for nondegenerate plasmas by Eqs.\rf{eps_Maxw},\rf{Zeffw0}, and\rf{kappa} are presented for different
ionic charges $Z=1,3,10$ as functions of the scaled frequency $\omega/\nu_e$ of the electromagnetic
radiation. The case of highly charged plasma ions with $Z=10$ is almost identical to the Lorentz model. The parameters $C$ and $s$ in Eq.\rf{kappa} were equal to $1$. For comparison the results of calculation by interpolation formula suggested by Brantov \emph{et.~al.}~\cite{Brantov08JETP} are also shown by dotted lines.
For considered long-wavelength perturbations ($k\to 0$) this interpolation formula consists of Eq.\rf{eps_MaxwL} with $\chii=1$ and the dimensionless
quantity $\xi_{\omega}$ is replaced by $\xi_{\omega} G_{Z}(\omega)$, where
\begin{equation}
G_{Z}(\omega ) = \dst\frac{1+C_0\xi_\omega/\gamma_\sigma(Z) }{1+C_0\xi_\omega}  ,\quad
C_0 = \frac{4}{15\sqrt{\pi}}(1+2i).
\label{G}
\end{equation}
Here the factor $\gamma_\sigma (Z)$ is given by Eq.~(\ref{sigma0}). In the limit $Z\gg 1$ the factor $\gamma_\sigma (Z)\to 1$ and therefore $G_{Z}(\omega)\to 1$, that gives the Lorentz model.

\begin{figure}[tbp]
\includegraphics[width=80mm]{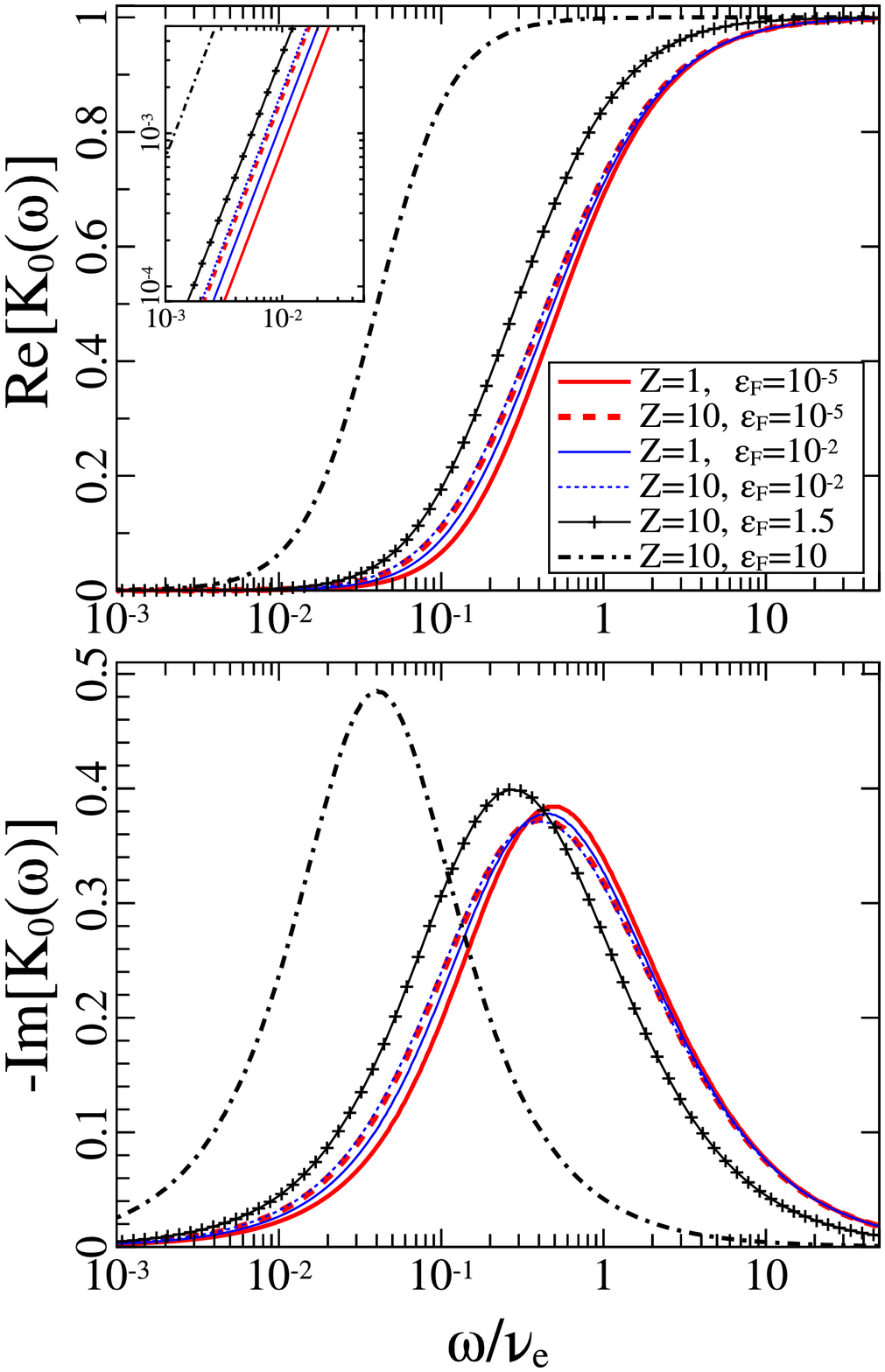}
\caption{(Color online) Real (top) and imaginary (with minus sign) (bottom) parts of $K_{0}(\omega )$,
calculated by Eqs.\rf{epsFermy},\rf{Zeffw0D}, and\rf{kappa} with $C=s=1$, for the degenerate electron
plasma with different ionic charges and different degeneracy parameters: $Z=1,\eF=10^{-5}$ (thick solid lines), $Z=10, \eF=10^{-5}$ (thick dashed lines),
$Z=1,\eF=10^{-2}$ (thin solid lines), $Z=10, \eF=10^{-2}$ (thin dashed lines),
$Z=10,\eF=1.5$ (marked lines), $Z=10,\eF=10$ (dash-dotted lines).}
\label{fig:K0_w_nu_Dg}
\end{figure}

It is seen that our results shown in\f{K0_w_nu} are very close to the interpolation results obtained in Ref.~\cite{Brantov08JETP}. The largest difference between both models occurs for imaginary part of the function $K_{0}(\omega )$ at $\omega/\nu_e \sim 0.5$ and $Z=1$ and the relative deviation is within 5\%. However, the interpolation formula of Ref.~\cite{Brantov08JETP} has itself the accuracy about 7\% compared to the more rigorous fully kinetic treatment~\cite{Brantov08JETP}.

It should be noted, that both models\rf{eps_Maxw},\rf{Zeffw0},\rf{kappa} and the interpolation formula suggested in Ref.~\cite{Brantov08JETP} lead to the correct asymptotic expressions for the permittivity in the low- and high-frequency limits, although interpolation formula~\cite{Brantov08JETP} does not satisfy the fundamental
property $\varepsilon(-\omega )=\varepsilon^{\ast }(\omega )$ and the Kramers-Kronig relations~\cite{LL_Electrodynamics}. This is because the function $G_{Z}(\omega)$ given by Eq.\rf{G}
does not satisfy the relation $G_{Z}(-\omega )=G^{\ast}_{Z}(\omega)$.
Unlike that, our model satisfies the equality $\varepsilon (-\omega )=\varepsilon^{\ast}(\omega )$ and the Kramers-Kronig relations.

It should be also emphasized, that the model presented here only weakly depends on the actual choice of the fitting parameters $C$ and $s$ in the expression\rf{kappa}. More specifically the results are only slightly changed in the interval $C,s\in [0.5;2]$.

In\f{K0_w_nu_Dg} the function $K_{0}(\omega )$, obtained by Eqs.\rf{epsFermy},\rf{Zeffw0D} and\rf{kappa},
is shown for the cases of partially degenerate plasmas with different degeneracy parameters $\eF =E_{\mathrm{F}}/T= 10^{-5}, 10^{-2}, 1.5, 10$ and different ionic charges $Z=1,10$.
The results for a weakly degenerate case with $\eF = 10^{-5}$ coincide for all $Z$ (thick solid and dashed lines in\f{K0_w_nu_Dg}) with ones calculated by Eqs.\rf{eps_Maxw},\rf{Zeffw0}, and\rf{kappa} obtained for nondegenerate plasma. For $Z\geqslant 10$ the results of calculations by Eqs.\rf{epsFermy},\rf{Zeffw0D} and\rf{kappa} are close to ones obtained for nondegenerate case if $\eF \lesssim 0.3$.

For $\eF \gtrsim 0.1$ the Spitzer factors\rf{sigma0D} for a degenerate plasma are very close to $1$. That is for moderately and highly degenerate plasmas the electron-electron collisions do not play significant role and $K_{0}(\omega )$ does not depend on $Z$.
For this case and for $\eF < 1$ (i.e. at $0.1\lesssim \eF < 1$) the dependence of $K_{0}(\omega )$ on the frequency is the same, as in the nondegenerate case with $Z\geqslant 10$.

As shown in\f{K0_w_nu_Dg} substantial difference between nondegenerate and degenerate regimes occurs at $E_\mathrm{F}/T \gtrsim 1$. For $E_\mathrm{F}/T \gg 1$ the difference is dramatic: the function $K_0(\omega)$ is shifted to the left along the $\omega/\nu_e$ axis while increasing $E_{\mathrm{F}}/T$. This is stipulated by the fact, that in accordance with Eq.\rf{epsFermyD}, the function $K_{0}(\omega)$ for a degenerate plasma depends on $\eta_{\omega }=\xi_\omega/\eF^{3/2}$, rather than on the parameter $\xi_\omega$ as in the nondegenerate case. This means that the displacement of the maximum of the function $K_0(\omega)$ along $\omega/\nu_e$  axis is proportional to $\eF^{3/2}$ for $\eF \gg 1$.
Therefore, to gain more insight we plot in\f{K0_eta_Dg} the function $K_{0}(\omega)$ versus the quantity $\eta_\omega^{-1}$, i.e. excluding the factor $\eF^{3/2}$ in the scaled frequency. One can easily see, that for $\eF \geqslant 5$ all curves are similar and centered near $\eta_\omega=1$ and for $\eF > 10$ one can use the Drude formula\rf{epsFermyHD} for calculation of the permittivity.

\begin{figure}[tbp]
\includegraphics[width=80mm]{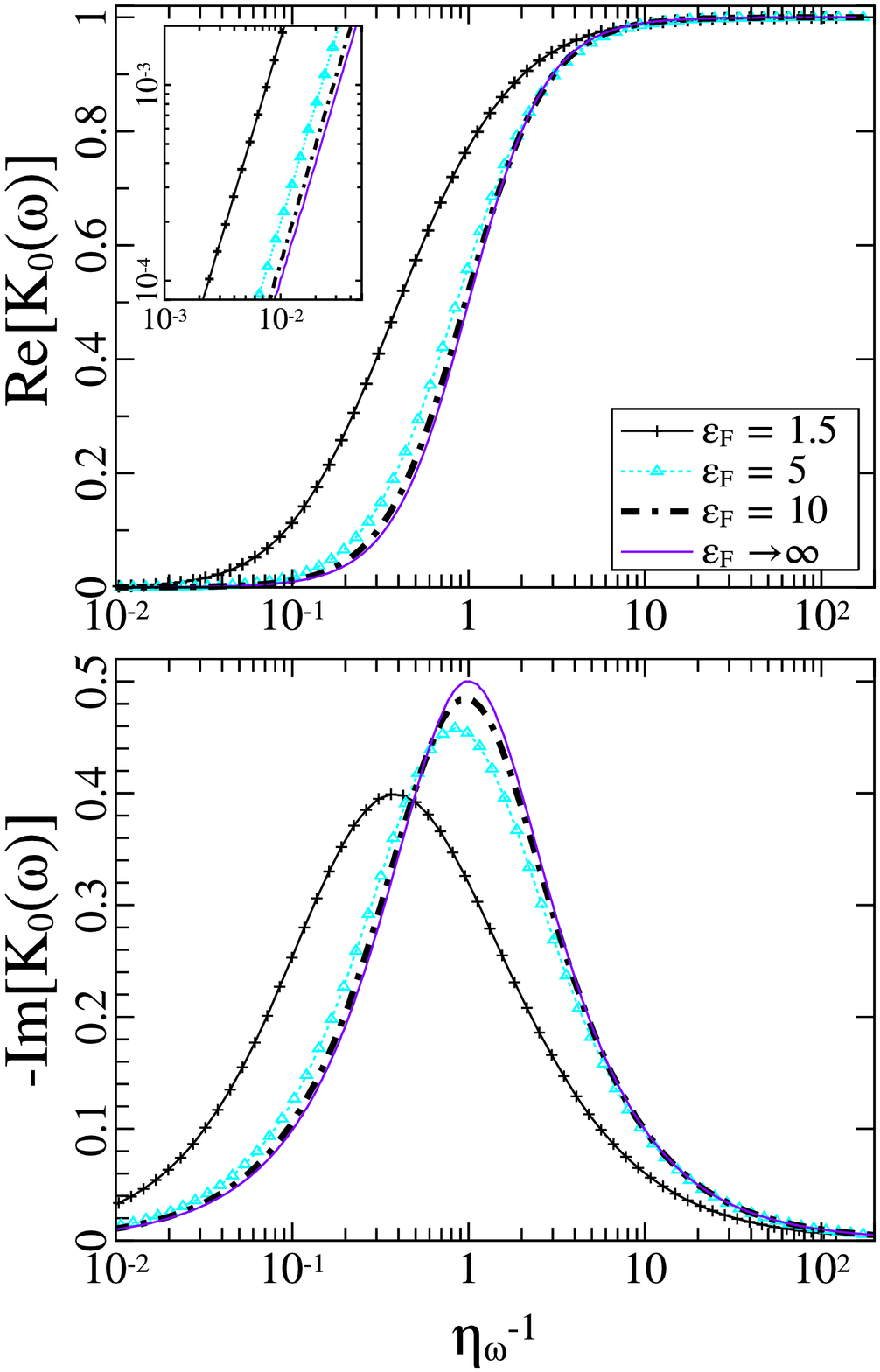}
\caption{(Color online) Real (top) and imaginary (with minus sign) (bottom) parts of $K_{0}(\omega)$
as functions of the parameter $\eta_\omega^{-1}$, calculated by Eqs.\rf{epsFermy},\rf{Zeffw0D},
and\rf{kappa} with $C=s=1$, for the degenerate electron plasma with $\eF=1.5$ (lines with crosses),
$\eF=5$ (lines with triangles),
$\eF=10$ (dash-dotted lines), and for $\eF\to\infty$ (solid lines). In the latter case the function $K_{0}(\omega)$ is given by
Eq.\rf{epsFermyD} which, however, in the limit $\eF\to\infty$ coincides with the Drude model\rf{epsFermyHD}. The results do not depend on the value of $Z$ (for $Z>1$ and $\eF > 1$).}
\label{fig:K0_eta_Dg}
\end{figure}

\section{Summary}
\label{sec:sum}

In this paper, we have obtained an analytical solution of the linearized Fokker-Planck kinetic equation with
a Landau collision integral and for a completely ionized, and unmagnetized electron plasma with an arbitrary
ionic charge. This solution accounts for both electron-ion collisions as well as the collisions of the subthermal
(cold) electrons with thermal ones. The latter collisions have been treated phenomenologically introducing
some parameter $\varkappa$ related to the relative contribution of the subthermal electrons to the friction
force and diffusion coefficient in velocity space [the limit $\varkappa \to 0$ corresponds to the vanishing
contribution of the electron-electron collisions].

Using the obtained solution of the Fokker-Planck
kinetic equation we have proposed an analytical model for the high-frequency ($\omega \gg kv_{\mathrm{th}}$) dielectric function of the collisional
electron plasma with an arbitrary ionic charge. More precisely the validity of the model is restricted to the
long-wavelength, high-frequency perturbations when $k^{-1}$ is a largest length scale of the problem with
$kv_{\mathrm{th}}\ll \omega$, $k\lambda_{ei}\ll 1$ and $k\lambda_{ee}\ll 1$, where $\lambda_{ei}$ and $\lambda_{ee}$
are the electron-ion and electron-electron mean free paths, respectively.

In our model the dielectric function
contains the contribution of the electron-electron collisions through unknown parameter $\varkappa (\omega)$
which has been treated as a function of the frequency $\omega$. Then $\varkappa (\omega)$ is adjusted considering
the low-frequency ($\omega \to 0$) limit of the dielectric function where it should agree with well-known expression
for the stationary electric conductivity. On the other hand, at high-frequencies ($\omega \to \infty$) it
behaves as $\varkappa (\omega) \to 0$ to fulfill the requirement of vanishing contribution of the electron-electron
collisions. One important feature of the outlined model is the possibility of generalization of the results to the
cases of a partially degenerate and/or strongly-coupled plasmas.
Making such generalization, we have assumed an additional limitation
$\hbar k^{2}/2m \ll \omega$ on the wavelength of the excitations.

In a further step we have considered a number of limiting cases: (a) limit of highly degenerate ($T\ll E_{\mathrm{F}}$) plasma,
(b) limit of low-frequencies, (c) limit of high-frequencies, (d) asymptotic behavior of the dielectric
function at large ionic charge, $Z\gg 1$, when our model coincides with the Lorentz plasma model derived either for
nondegenerate~\cite{LL_PhysKin} or partially degenerate plasmas~\cite{Lee84PhysFluids,Kostenko08GSI}. These limiting cases facilitate
the systematic comparison of our analytical results with the previous theoretical models.

In particular, the present
model has been compared both analytically and numerically with the interpolation formula suggested by Brantov~\emph{et.~al.}
\cite{Brantov08JETP}. It has been demonstrated that our results agree satisfactory well with ones obtained in Ref.~\cite{Brantov08JETP}
showing relative deviations less than 5\% in an unfavorable case of lowest ionic charge $Z=1$. It should be noted, however,
that the interpolation formula by Brantov~\emph{et.~al.} has the accuracy about 7\% compared to the more rigorous fully
kinetic treatment of Ref.~\cite{Brantov08JETP}.

As the main goal of this paper we suggest a simple but more advanced analytical model for calculations of the dielectric
function and related quantities in a wide range of parameters which is appropriate for modeling many experiments with
laser-matter interactions. In addition, further improvement of the present model can be achieved by considering the
spatial inhomogeneity of the perturbations (i.e.~finite wavelengths $k^{-1}$) in the Fokker-Planck kinetic equation~(\ref{f}).
This can be done using the method of Ref.~\cite{Brantov08JETP} for the solution of the kinetic equation and, for treating
the electron-electron collisions, following the same steps that led to the approximate coefficients~(\ref{Dij2})
and (\ref{Fi2}). Systematic investigation of this problem is left for future work.

\begin{acknowledgments}
The work of H.B.N. and H.H.M. was supported by the State
Committee of Science of the Armenian Ministry of Higher Education and
Science (Project No.~13-1C200). The work of M.E.V. and N.E.A. was supported
in part by the programs on fundamental research of the Russian Academy of Sciences.
\end{acknowledgments}

\bibliography{pre2014}

\end{document}